\documentclass[a4paper,11pt]{article}
\usepackage{pos}
\usepackage{subfigure}
\usepackage{url}
\usepackage{siunitx}
\title{Detecting ultra-high-energy cosmic rays with prototypes of the Fluorescence detector Array of Single-pixel Telescopes (FAST) in both hemispheres}
 \ShortTitle{Detecting UHECRs with prototypes of the FAST in both hemispheres}
\newcommand{\xmax}{X_\mathrm{max}}

\author*[a]{Shunsuke~Sakurai}
\author[b]{Justin~Albury}
\author[b]{Jose~Bellido}
\author[a]{Fraser~Bradfield}
\author[c]{Ladislav~Chytka}
\author[d]{John~Farmer}
\author[a]{Toshihiro~Fujii}
\author[e]{Petr~Hamal}
\author[e]{Pavel~Horvath}
\author[e]{Miroslav~Hrabovsky}
\author[e]{Vlastimil~Jilek}
\author[e]{Jakub~Kmec}
\author[e]{Jiri~Kvita}
\author[d]{Max~Malacari}
\author[c]{Dusan~Mandat}
\author[f]{Massimo~Mastrodicasa}
\author[g]{John~N.~Matthews}
\author[e]{Stanislav~Michal}
\author[h]{Hiromu~Nagasawa}
\author[h]{Hiroki~Namba}
\author[e]{Libor~Nozka}
\author[c]{Miroslav~Palatka}
\author[c]{Miroslav~Pech}
\author[d]{Paolo~Privitera}
\author[f]{Francesco~Salamida}
\author[c]{Petr~Schovanek}
\author[d]{Radomir~Smida}
\author[e]{Daniel~Stanik}
\author[e]{Zuzana~Svozilikova}
\author[i]{Akimichi~Taketa}
\author[h]{Kenta~Terauchi}
\author[g]{Stan~B.~Thomas}
\author[c,e]{Petr~Travnicek}
\author[c]{Martin~Vacula}

\affiliation[a]{Graduate School of Science, Osaka Metropolitan University, Sumiyoshi-ku, Osaka, Japan}
\affiliation[b]{Department of Physics, University of Adelaide, Adelaide, S.A., Australia}
\affiliation[c]{Institute of Physics of the Academy of Sciences of the Czech Republic, Prague, Czech Republic}
\affiliation[d]{Kavli Institute for Cosmological Physics, University of Chicago, Chicago, IL, USA}
\affiliation[e]{Joint Laboratory of Optics of PU and IF of CAS, Palacky University, Olomouc, Czech Republic}
\affiliation[f]{Department of Physical and Chemical Sciences, University of L’Aquila and INFN LNGS}
\affiliation[g]{High Energy Astrophysics Institute and Department of Physics and Astronomy, University of Utah, Salt Lake City, UT, USA}
\affiliation[h]{Graduate School of Science, Kyoto University, Sakyo-ku, Kyoto, Japan}
\affiliation[i]{Earthquake Research Institute, University of Tokyo, Bunkyo-ku, Tokyo, Japan}
\emailAdd{ssakurai@omu.ac.jp}

\abstract{
Ultra-high energy cosmic rays (UHECRs), whose energy are beyond \SI{e18}{eV}, are the most energetic particles we have ever detected.
The latest results seem to indicate a heavier composition at the highest energies, complicating the search for their origins.
Due to the limited number of UHECR events, 
we need to build an instrument with an order of magnitude larger effective-exposure to collect UHECRs in future decades.
The Fluorescence detector Array of Single-pixel Telescopes (FAST) is a proposed low-cost, easily deployable UHECR detector suitable for a future ground array.
It is essential to validate the telescope design and autonomous observational techniques using prototypes located in both hemispheres.
Here we report on the current status of observations, recent performance results of prototypes, and developments towards a future mini-array.
}

\ConferenceLogo{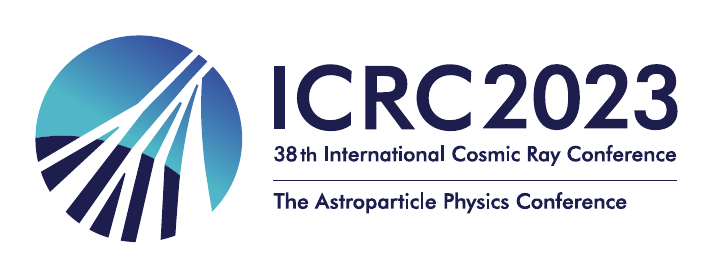}

\FullConference{
38th International Cosmic Ray Conference (ICRC2023)\\
  26 July - 3 August, 2023\\
  Nagoya, Japan}

\begin{document}
\maketitle

\section{Recent progress and open questions of ultra-high energy cosmic rays}
\begin{figure}[ht]
    \centering
    \includegraphics[width=0.9\textwidth]{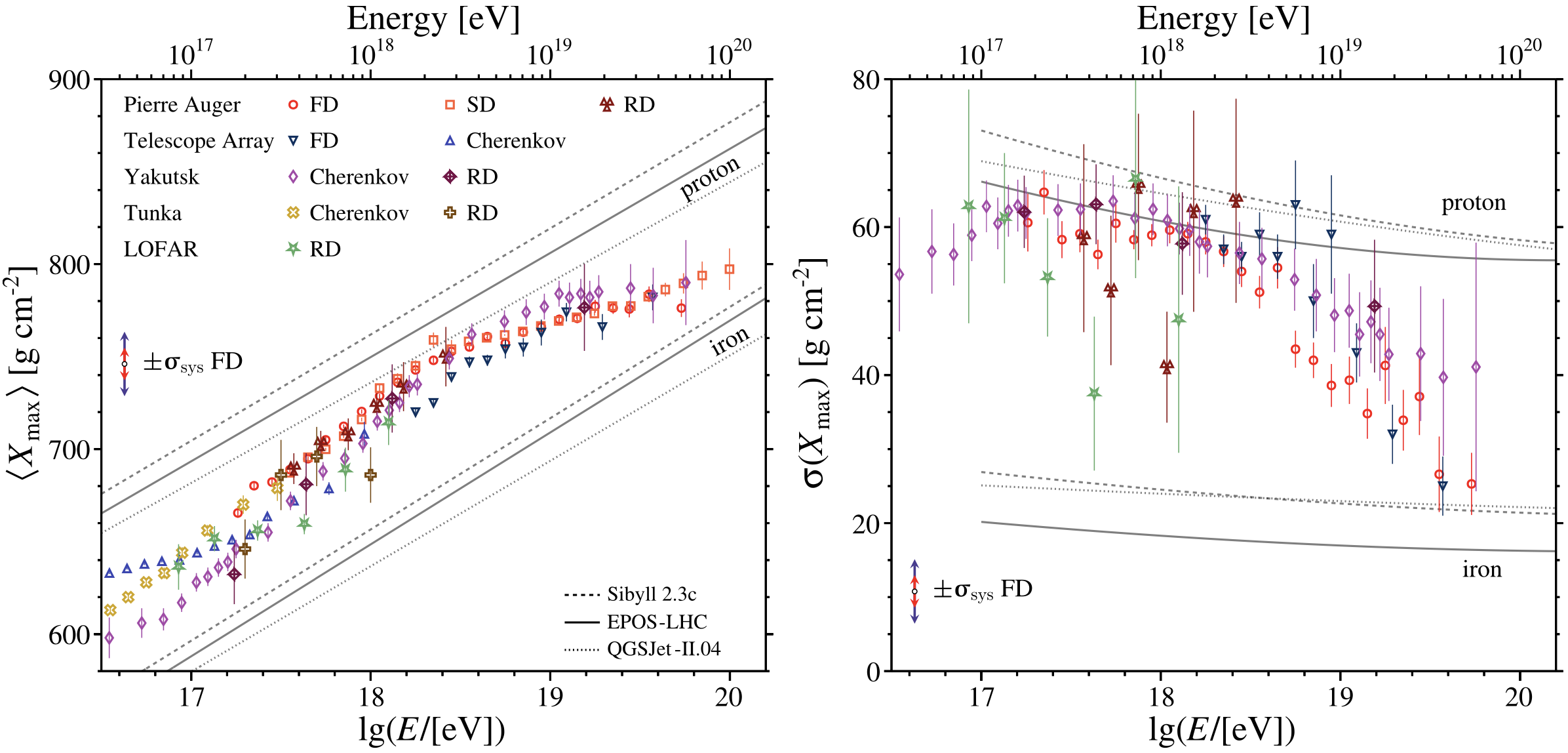}
    \caption{Depth of air-shower profile maximum ($\xmax$) measured by several experiments. The left shows mean $\xmax$ and the right is standard deviation of $\xmax$. Results imply UHECR mass getting heavier at the highest energies~\cite{COLEMAN2023102819:snowmass}.}
    \label{fig:enter-label}
\end{figure}

Ultra-high energy cosmic rays (UHECRs) are the most energetic particles ever detected at Earth. 
They are thought to be related to extremely energetic astrophysical phenomena. However, there is still no clear indication of their origins. 
Two leading experiments, Pierre Auger Observatory (Auger)~\cite{bib:auger} and Telescope Array experiment (TA)~\cite{bib:tafd, bib:tasd}, have been conducting cosmic-ray observations with combinations of surface detector arrays (SD) and fluorescence detectors (FD).
After more than 15 years observations with Auger and TA, there is evidence for possible source candidates and a heavier mass composition at the highest energies~\cite{COLEMAN2023102819:snowmass}.
Figure~\ref{fig:enter-label} shows the depth of air-shower profile maximum ($\xmax$) measured by several experiments~\cite{COLEMAN2023102819:snowmass}. 
Since $\xmax$ is a proxy for measuring primary particle composition,
we can infer that the primary composition becomes proton-dominated at above \SI{e18}{eV} and gradually changes to heavier nuclei at the highest energies. 
This feature is important to understand when searching for UHECR origins, because heavier nuclei are strongly deflected by magnetic fields during thier propagation.

The flux of UHECRs follows a power-law function $\sim E^{-3}$, where $E$ is the primary energy of UHECR, so the flux of particles at the highest energies is very low. In addition,
interactions with the cosmic microwave background via the Greisen-Zatsepin-Kuzmin process further suppress this flux~\cite{bib:gzk1,bib:gzk2}.
Therefore, collecting sufficient statistics for source identification requires a huge effective detection area.
To achieve this, low-cost, easily deployable and maintenance-free autonomous detectors are essential for the next generation observatories.
Several projects are ongoing to realize the next generation of UHECR astronomy~\cite{Olinto_2021:poemma,grand:2022,Tameda:2021GM,Brown:2021tf,Horandel:2021prj}.

\section{Fluorescence detector Array of Single-pixel Telescopes (FAST)}
Fluorescence detector Array of Single-pixel Telescopes (FAST)\footnote{\url{https://www.fast-project.org}} is one of the next-generation UHECR projects, which aims to observe UHECRs via atmospheric fluorescence light emitted by extensive air showers~\cite{bib:fast,Malacari:2019uqw}.
FAST is designed to achieve \SI{e6}{km^2.sr.yr} exposure with 500 stations spread over \SI{150000}{km^2}, which is an order of magnitude higher than that of Auger and TA.
Each station will consist of 12 FAST telescopes.
A single FAST telescope has a \SI{1.6}{m} diameter segmented mirror and four \SI{20}{cm} photo-multiplier tubes (PMTs) at the focal plane of the mirror~\cite{bib:fast_optics}.  

\begin{figure}[ht]
    \centering
    \subfigure[Expected sensitivity]{\includegraphics[width=0.52\textwidth]{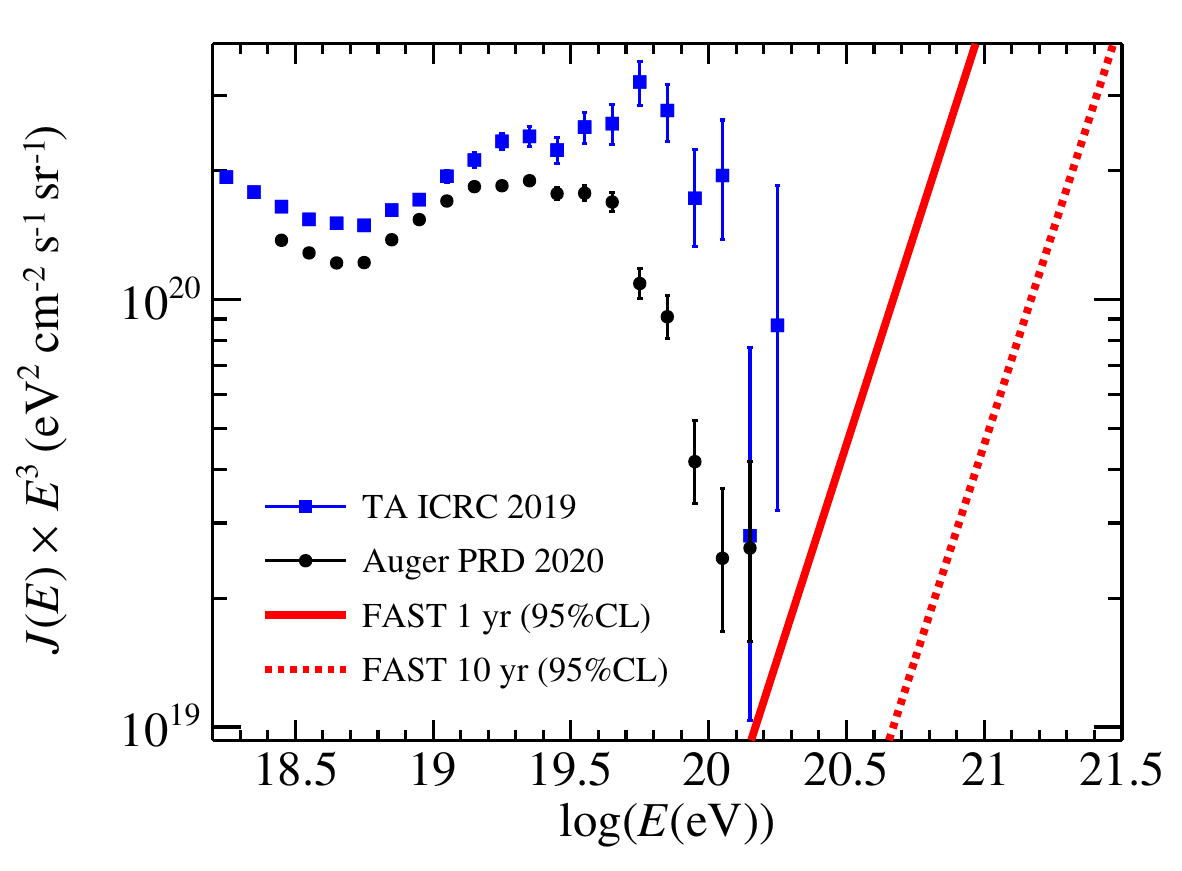}}
    \subfigure[Reconstructed $\xmax$ distribution]
    {\includegraphics[width=0.46\textwidth]{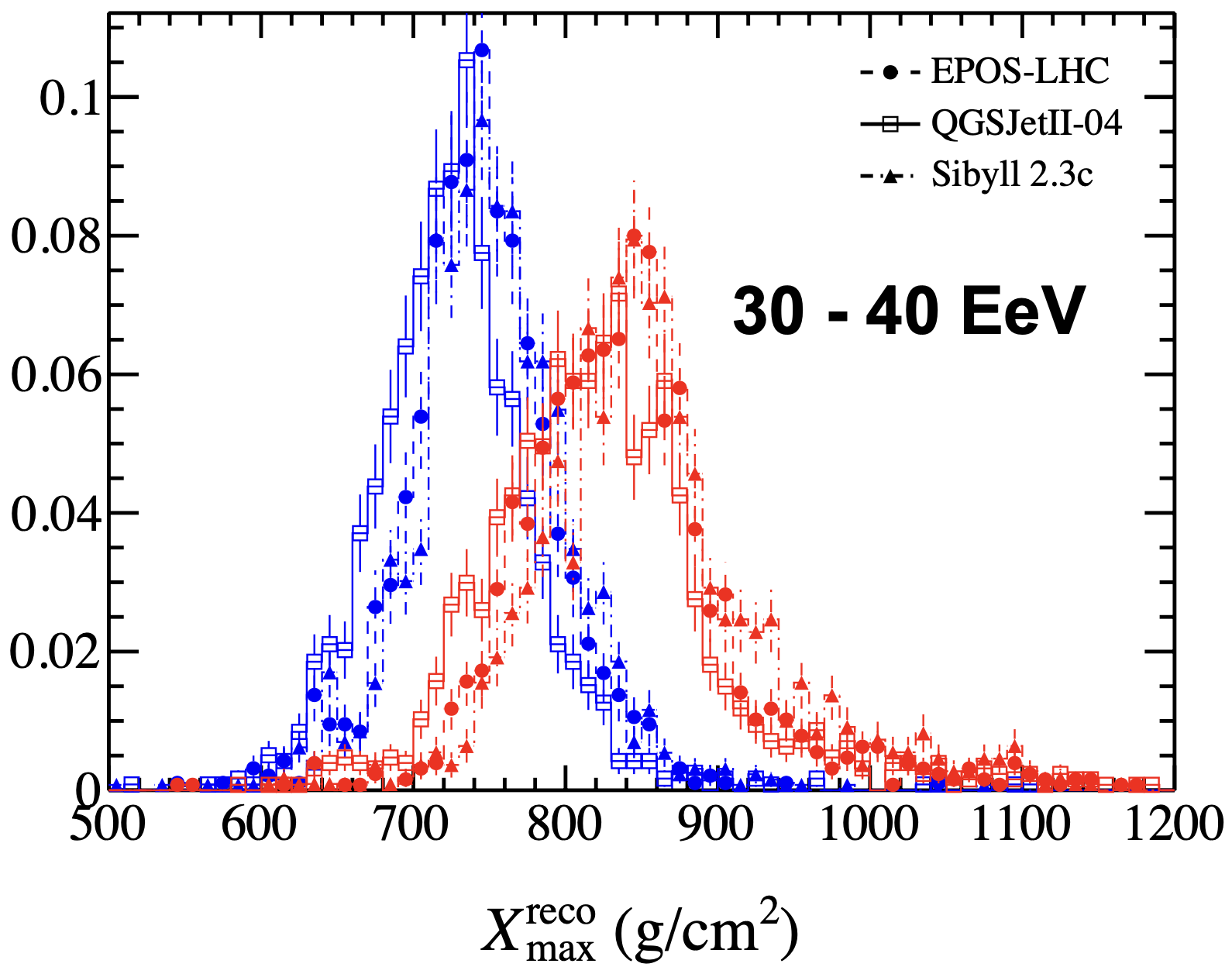}}
    \caption{Simulated performance with full-scale FAST array compared to the current observatories. Left: Expected FAST sensitivity in 1 year and 10 years. Right: Reconstructed $\xmax$ distribution for proton (red) and iron (blue) primaries with three hadronic interaction models~\cite{fujii:uhecr22}.}
    \label{fig:enter-labelB}
\end{figure}

FAST focuses on the detection of UHECRs with energies above \SI{3e19}{eV}. Ten years of operation will enable FAST to investigate the energy spectrum beyond \SI{3e20}{eV}, as shown in Figure~\ref{fig:enter-labelB}(a).
Applying neural network reconstruction to the Monte-Carlo simulations, we find the energy and $\xmax$ resolutions to be $\SI{8}{\percent}$ and $\SI{30}{g.cm^{-2}}$ at \SI{4e19}{eV}~\cite{fujii:uhecr22}.
The reconstructed $\xmax$ distribution for simulated proton and iron primaries with three hadronic interaction models are shown in Figure~\ref{fig:enter-labelB}(b). 

\begin{figure}[ht]
    \centering
    \includegraphics[width=0.9\textwidth]{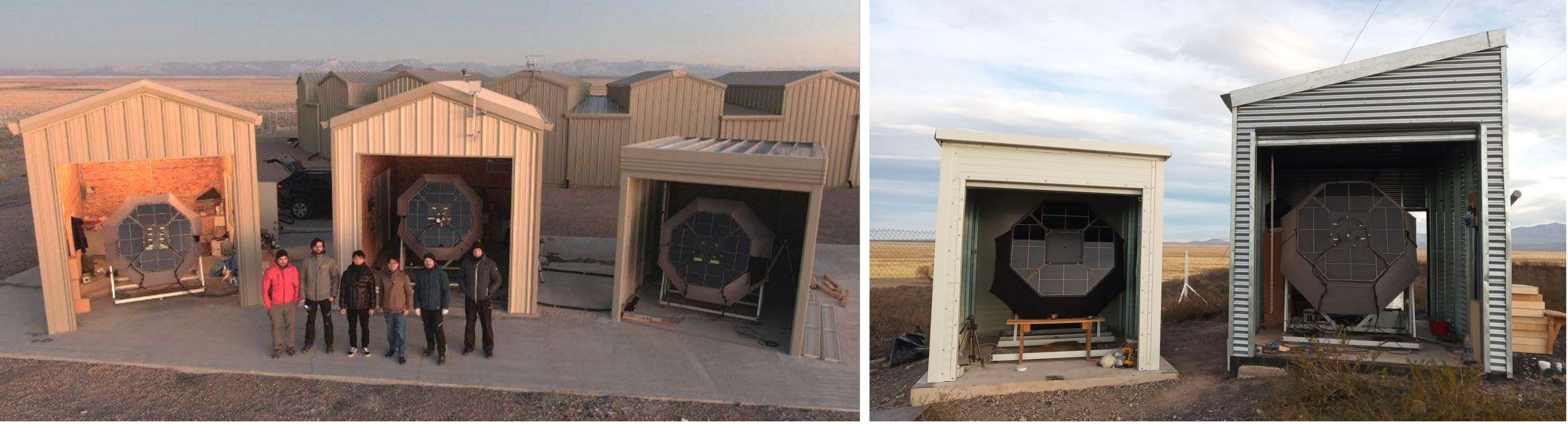}
    \caption{FAST prototypes in the Auger and TA site. Left: Three prototypes at TA are in operation. Right: Two prototypes at Auger. The second telescope (left in the picture) is not in operation yet~\cite{fujii:uhecr22}.}
    \label{fig:enter-labelC}
\end{figure}

Compared to current-generation FDs, the design of FAST is simplified to meet the financial and deployability requirements.
Therefore, the observation and analysis techniques need to be validated and confirmed before proceeding to mass production.
Observations and performance evaluations are being carried out using prototype telescopes located at the Auger and TA sites.
Three prototypes have been deployed at the TA Black Rock Mesa site (FAST@TA), and two prototypes at the Auger Los Leones site (FAST@Auger), as shown in Figure~\ref{fig:enter-labelC}.
Here we report the current status of observations conducted thus far as well as the performance evaluation from the data collected. Furthermore, we plan to build a small prototype array of FAST (FAST mini array) with \SI{20}{km} spacing at the Auger site.
Developments regarding the FAST mini array are also reported.

\section{Observation status and performance estimation of FAST prototypes}
\begin{figure}[ht]
    \centering
    \subfigure[FAST event display]{\includegraphics[width=0.52\textwidth]{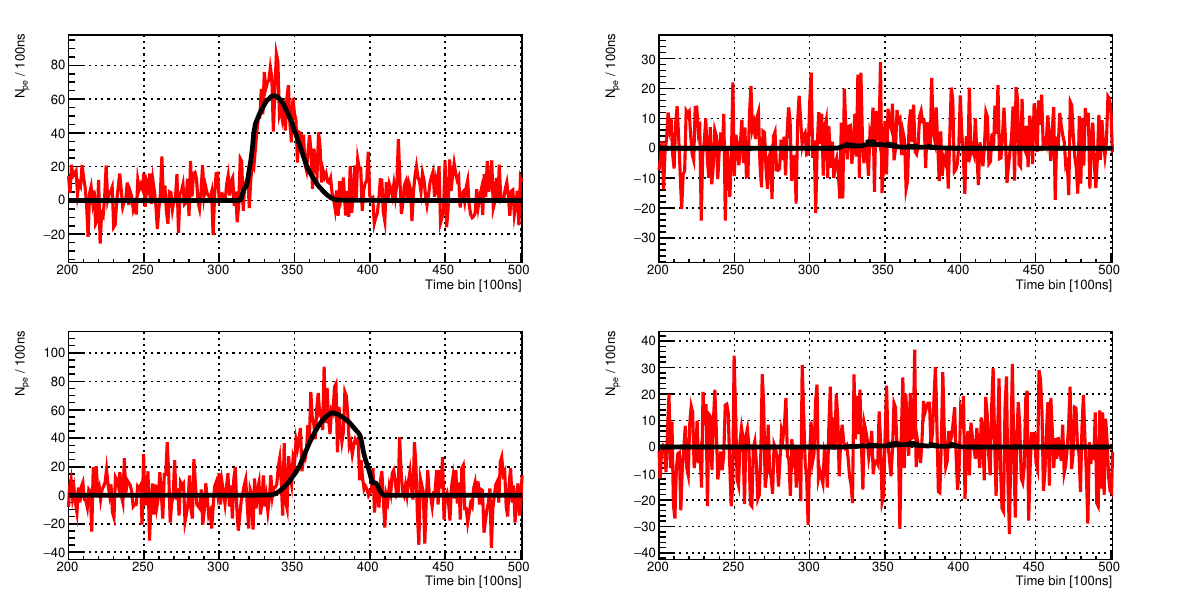}}
    \subfigure[TA event display]
    {\includegraphics[width=0.46\textwidth]{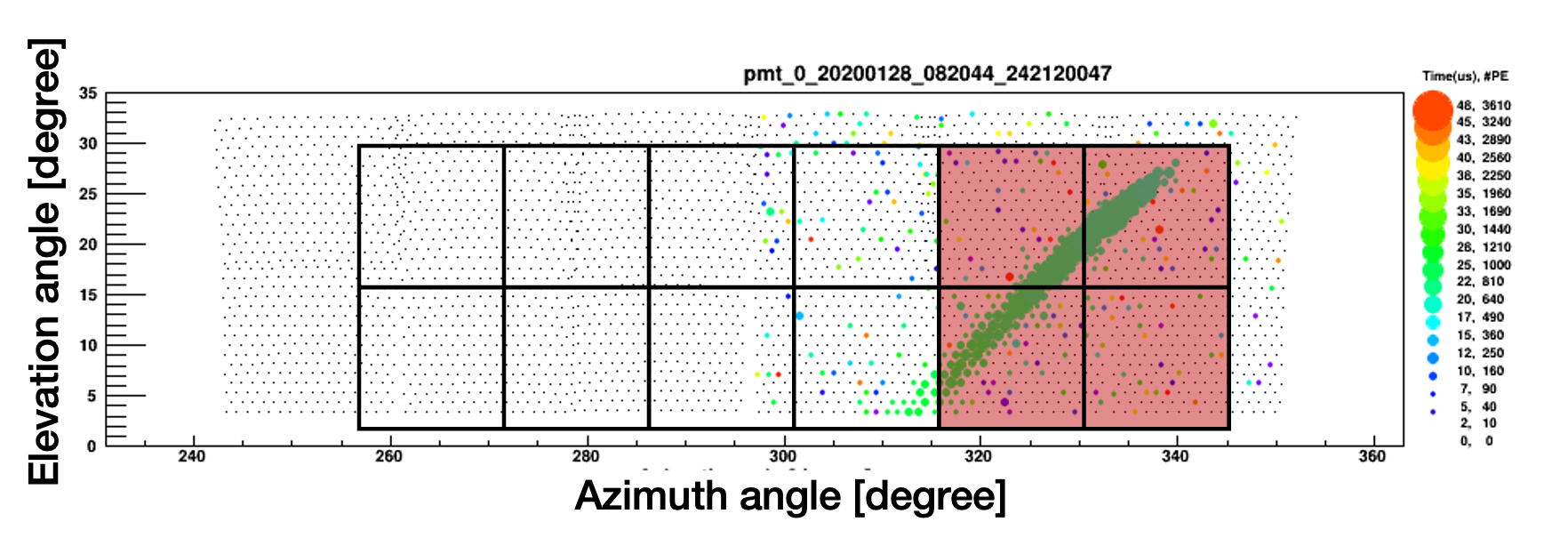}}
    \caption{An example of the coincidence UHECR event measured with FAST@TA and TA FD. Left: Comparison of observed data (red) and the best fit model (black). Right: TA FD event display of the same event, overlapping the field of view of individual PMTs of FAST@TA. Colors show relative timings and the number of photo-electrons recorded by TA FD PMTs.}
    \label{fig:enter-labelD}
\end{figure}

FAST@TA has a field of view (FoV) of $\SI{30}{\degree} \times \SI{120}{\degree}$, and overlapping the TA FD FoV.
FAST@TA has observed cosmic rays since 2016 with an external trigger from the TA FD, except for a break due to COVID-19.
The total observation time is 637 hours. 
The FAST data analysis is based on the maximum likelihood technique, which compares observed waveforms with expected ones from simulations.
The simulated reconstruction performance of FAST@TA is reported in another ICRC contribution~\cite{Bradfiled:2023Li}. 
Figure~\ref{fig:enter-labelD} shows a UHECR event detected by FAST@TA in January 2019. The event was also detected by TA FD. The best fit model agrees well with observed data. The reconstructed energy and $\xmax$ by FAST@TA are \SI{1.7e18}{eV} and \SI{816}{g.cm^{-2}}. We have reported 59 significant events between March 2018 and October 2019 in the previous ICRC~\cite{Fujii:2021ab}. We found additional 20 UHECR candidates since 2019. They are being analyzed and compared with the TA FD analysis.

\begin{figure}[ht]
    \centering
    \subfigure[FAST event display]{\includegraphics[width=0.46\textwidth]{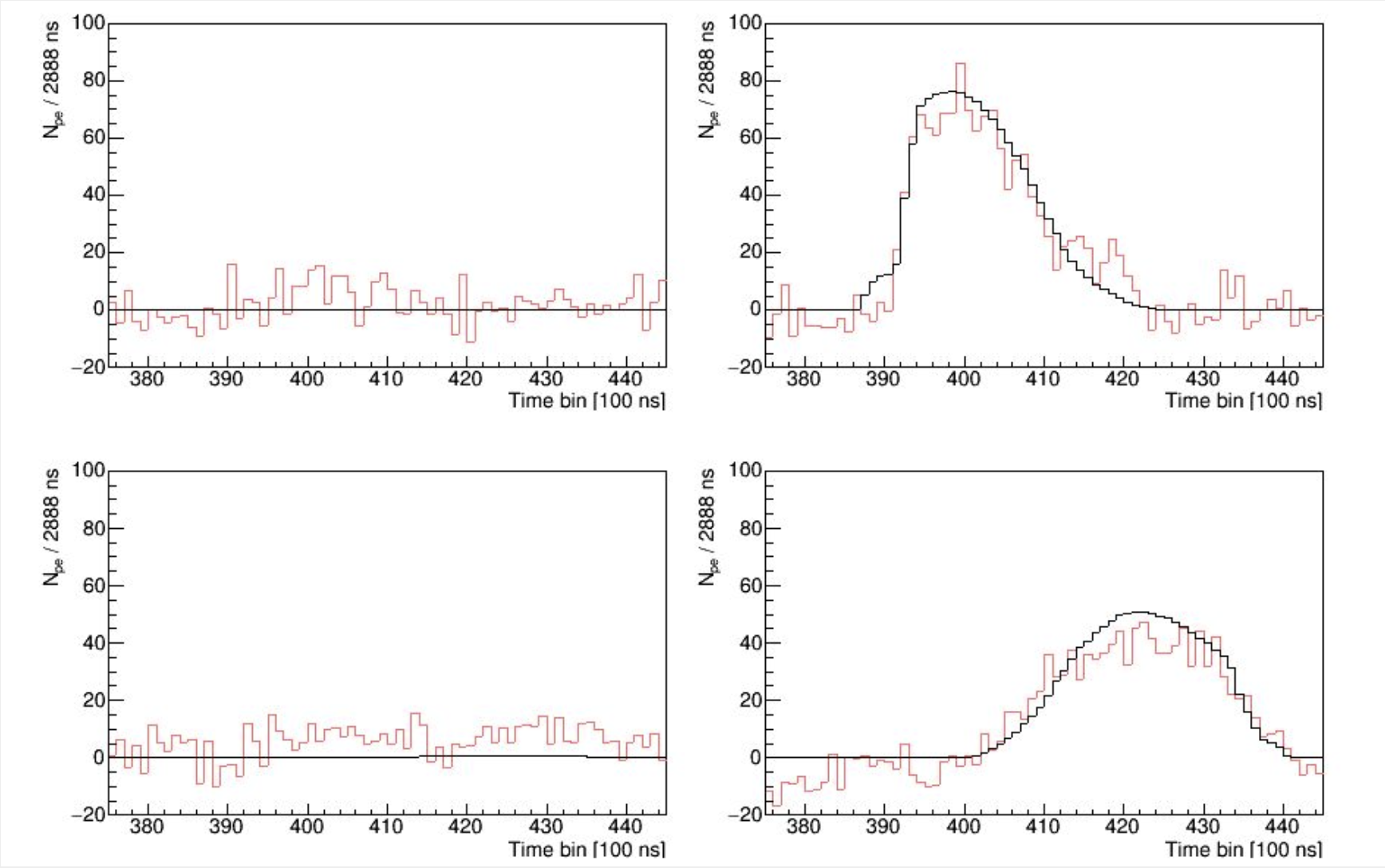}}
    \subfigure[Auger event display ]
    {\includegraphics[width=0.52\textwidth]{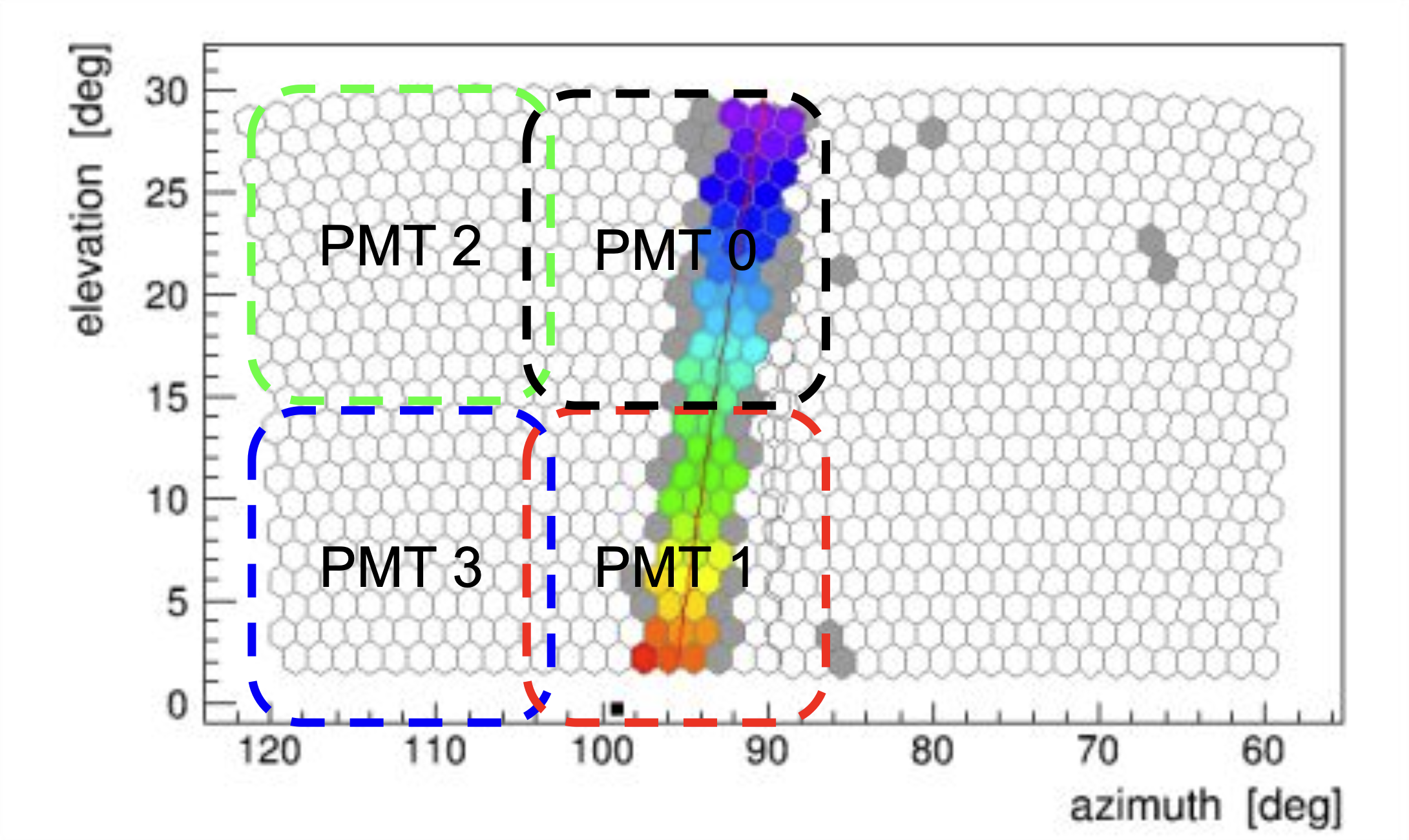}}
    \caption{An example of the coincidence UHECR event measured with FAST@Auger and Auger FD. Left: Comparison of observed data (red) and the best fit model (black). Right: Auger FD event display of the same event, overlapping the field of view of individual PMTs of FAST@Auger.}
    \label{fig:enter-labelE}
\end{figure}

One prototype telescope at FAST@Auger has been in regular operation since 2019.
A second prototype was installed in 2022 and is waiting for the installation of electronics later this year.
Analysis comparing FAST@Auger data with Auger FD data is being conducted.
Figure~\ref{fig:enter-labelE} shows an example of the coincidence UHECR event measured with both FAST@Auger and Auger FD.
The reconstructed energy and $\xmax$ by FAST@Auger are \SI{1.12e18}{eV} and \SI{632}{g.cm^{-2}} .

\section{Developments for the FAST mini array}
FAST will deploy 500 stations over an unprecedented area, and is expected to observe UHECRs for more than 20 years.
Stand-alone observations with autonomous telescopes are crucial.
We will build the FAST mini array with six FAST prototypes at the Auger observatory to validate our concept.
Developments and optimizations of the prototype telescope design are ongoing.
Here we report on new electronics, a new enclosure, low-cost mirrors and upgraded PMTs.

\begin{figure}[ht]
    \centering
    \subfigure[New custom electronics]{\includegraphics[width=0.3\textwidth]{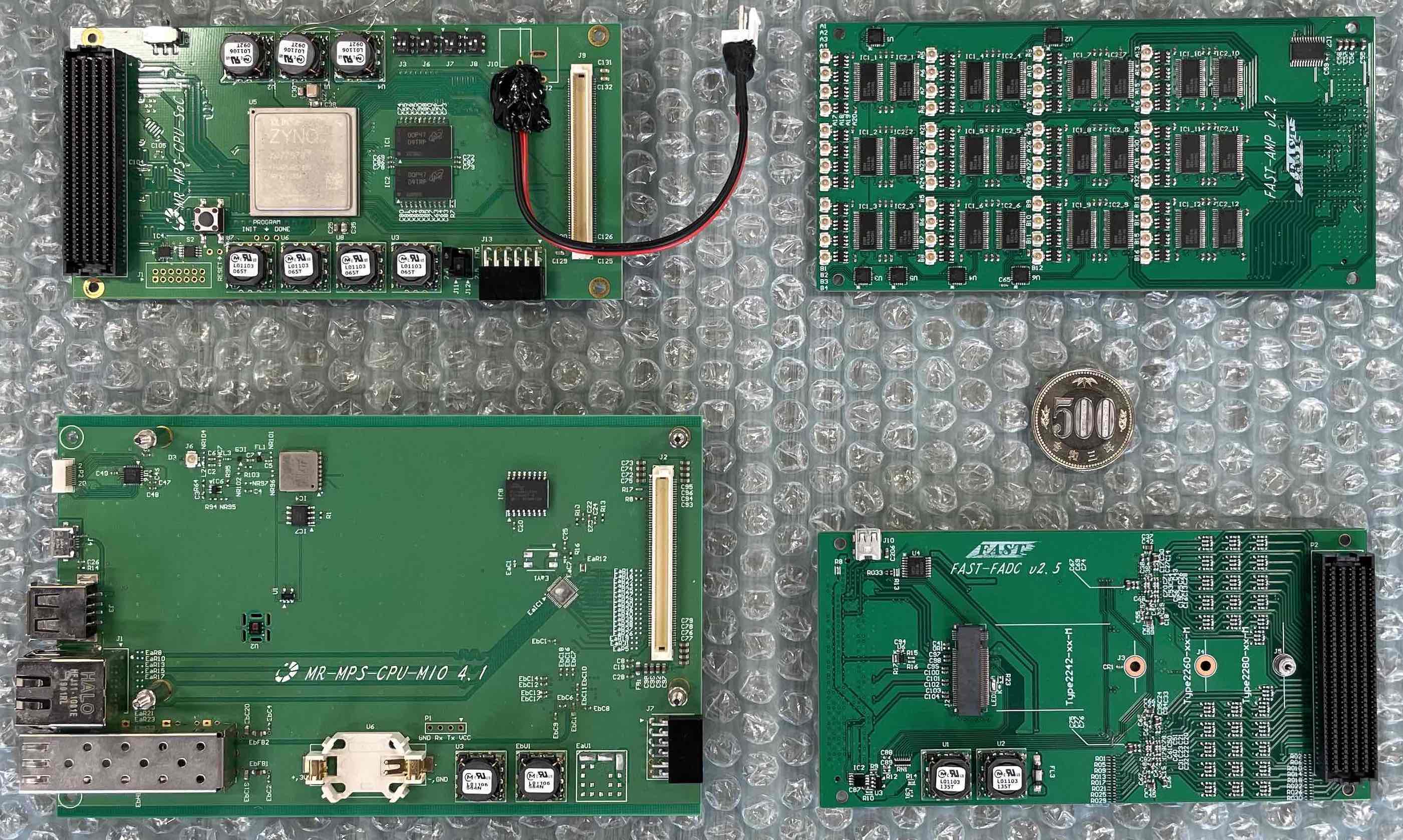}} 
    \subfigure[Onsite test]{\includegraphics[width=0.25\textwidth]{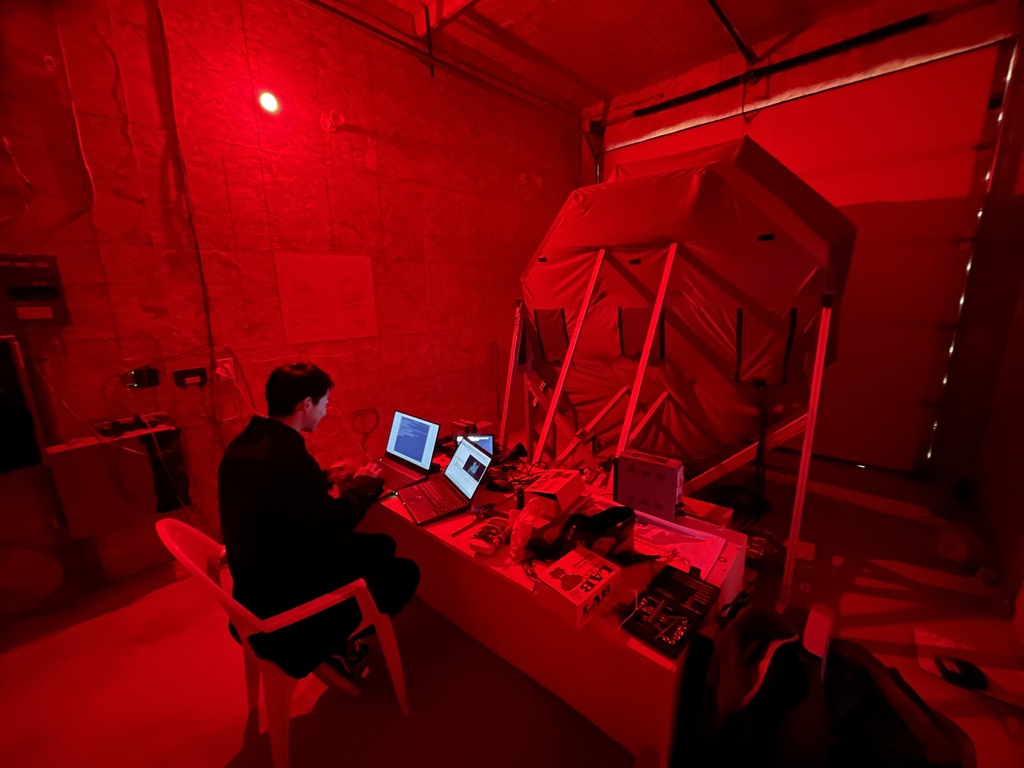}} 
    \subfigure[A signal expected from an extensive air shower]{\includegraphics[width=0.3\textwidth]{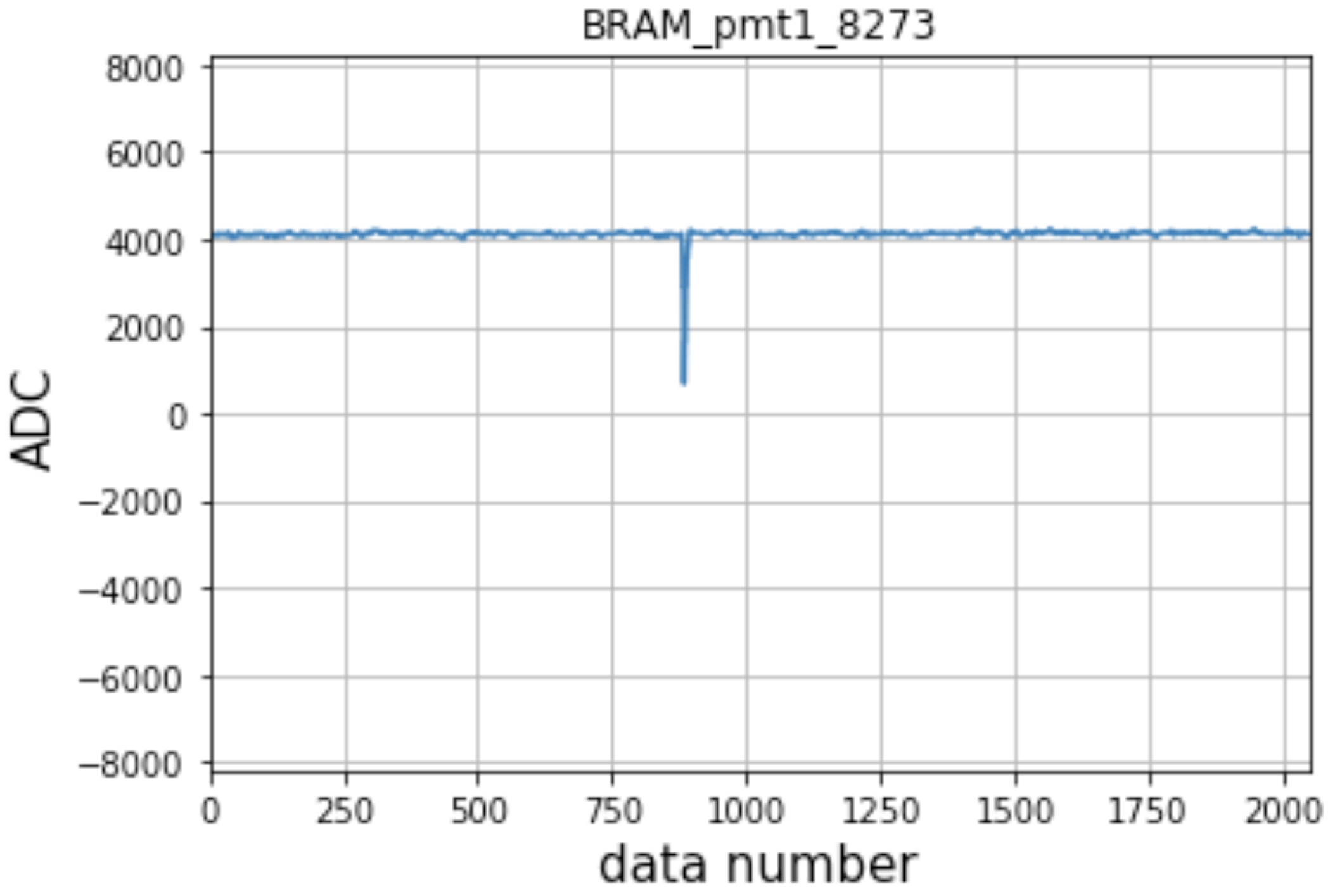}} 
    \caption{New custom electronics and onsite test. Left: prototype circuit boards of new custom electronics, Middle: Onsite test, Right: Recorded signal with the custom electronics triggering by external triggers. Signal seems to be from extensive air showers.}
    \label{fig:enter-labelF}
\end{figure}

The current FAST electronics is based on off-the-shelf commercial electronics that receives external triggers to record PMT waveforms.
New custom electronics will enable us to issue internal triggers and perform data taking independently.
The prototype custom electronics are being tested in Osaka Metroplotian University as shown in Figure~\ref{fig:enter-labelF}(a).

The firmware development for stand-alone observations is in progress.
Onsite tests of the prototype electronics with external triggers were carried out in November 2022.
Figure~\ref{fig:enter-labelF}(b) shows the picture of the test performed at the TA site.
We were able to trigger on a signal expected from an extensive air shower as shown in Figure~\ref{fig:enter-labelF}(c). Time coincidence signals were confirmed in other PMTs.

\begin{figure}[ht]
    \centering
    \subfigure[A conceptual design of new FAST]{\includegraphics[width=0.27\textwidth]{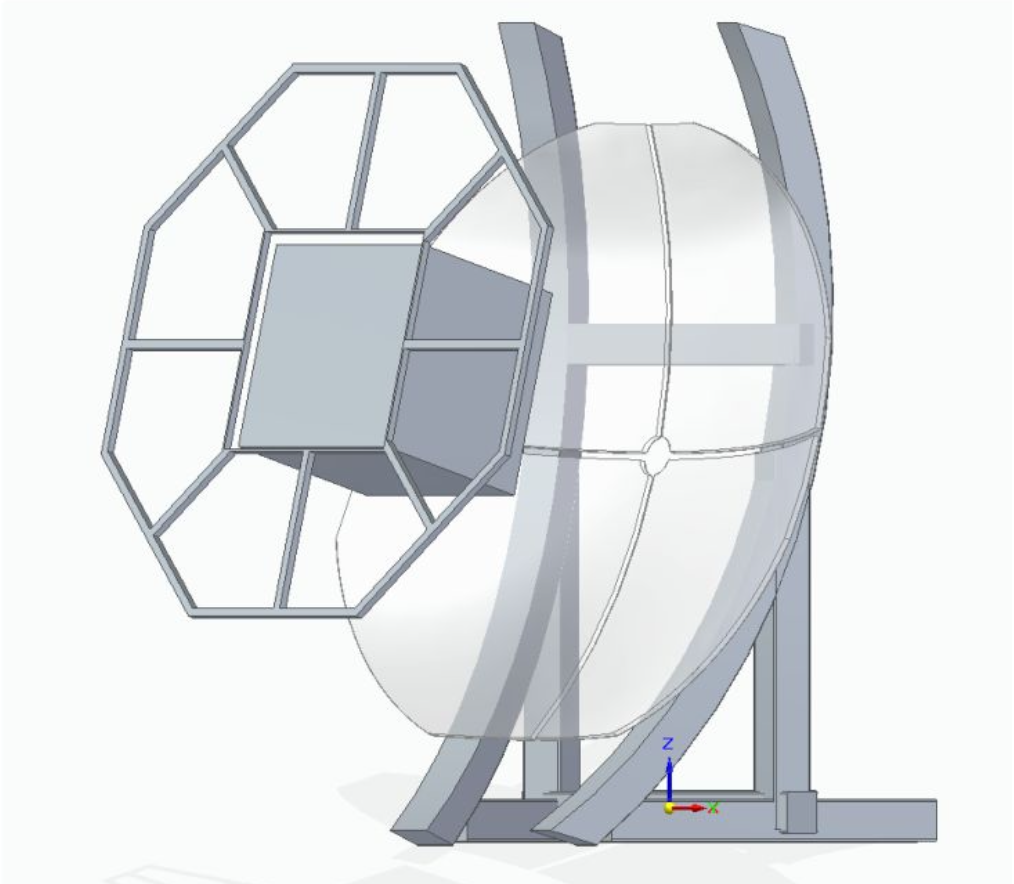}}
    \subfigure[An enclosure for future prototypes]{\includegraphics[width=0.33\textwidth]{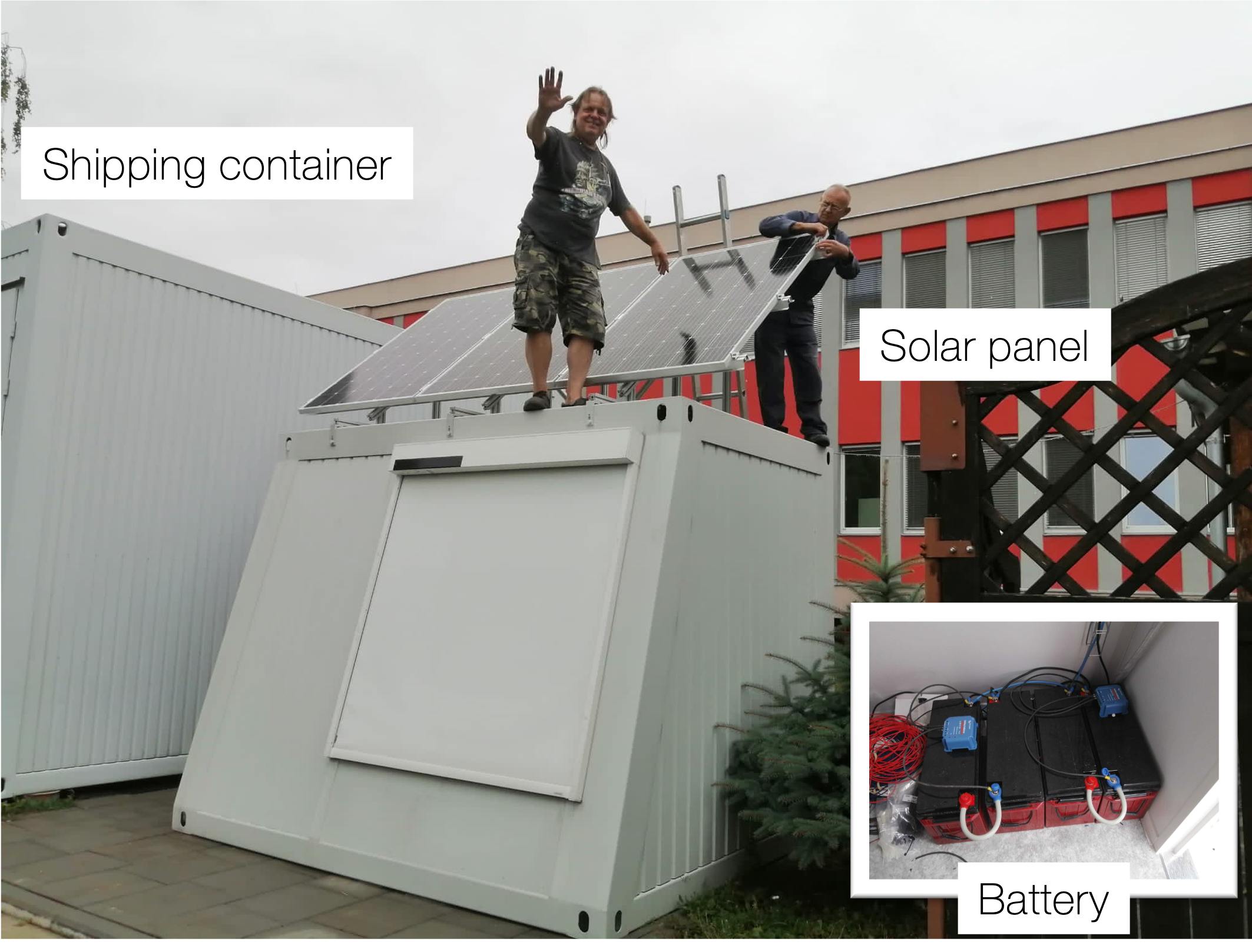}}
    \subfigure[A comparison of angular efficiencies
    ]{\includegraphics[width=0.29\textwidth]{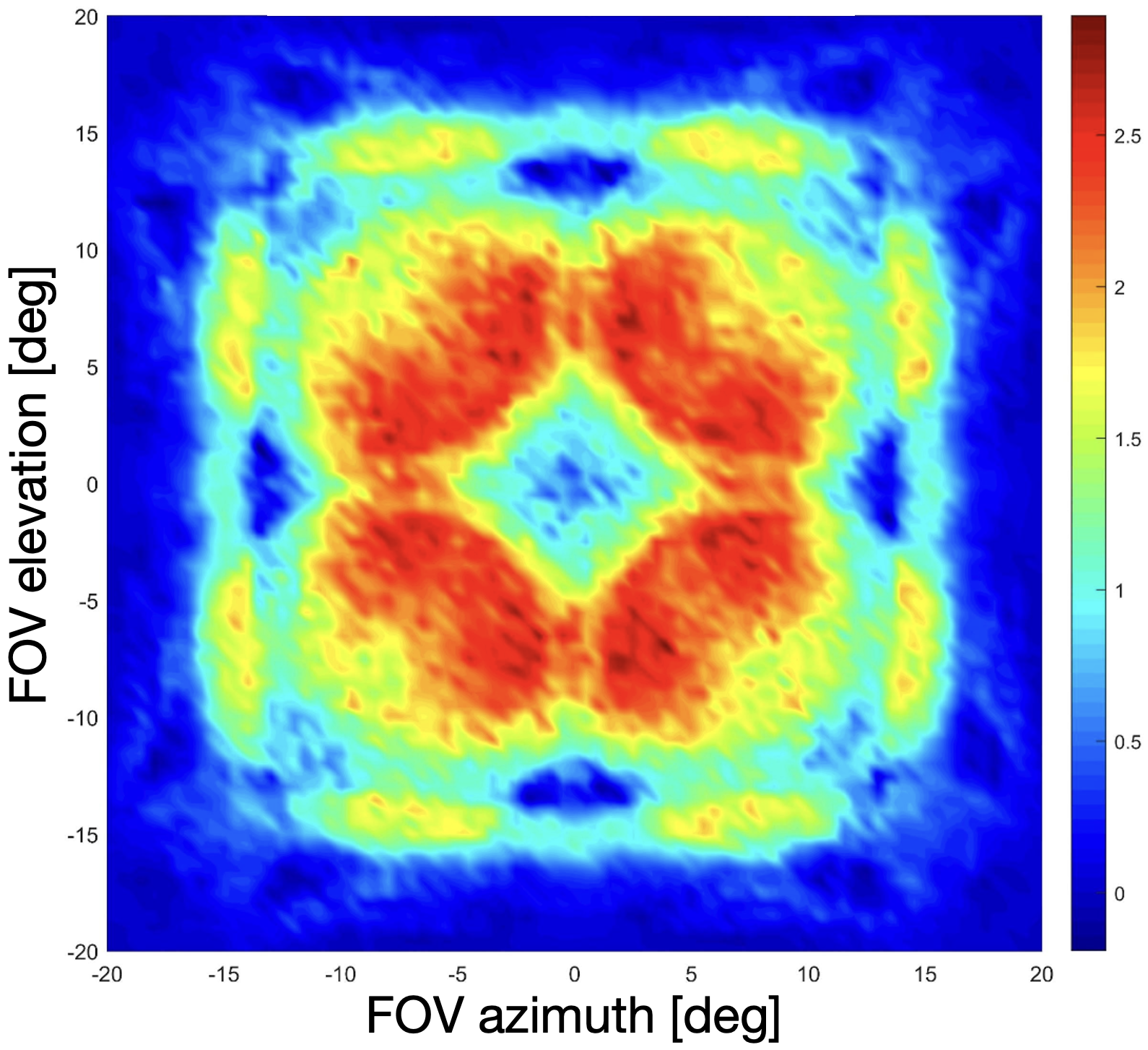}}
    \caption{Developments for the FAST mini array. Optimization of telescope design is being discussed and some components are already produced for the next prototypes.}
    \label{fig:enter-labelG}
\end{figure}

Optimizations of the telescope design and its enclosure to further reduce costs are continuing.
Figure~\ref{fig:enter-labelG}(a) shows a new telescope design with reduced weight and a mirror with fewer segments.
The new enclosure has been constructed in Olomouc, Czech Republic as shown in Figure~\ref{fig:enter-labelG}(b).
The enclosure was designed to fulfill deployabilty requirements.
A solar panel is equipped on the enclosure for stand-alone observation.
Batteries are charged during daytime through the solar panel and used for the observation. 
Field tests will be performed at Ondrejov, Czech Republic.

The new mirror design and simplified manufacturing process will speed up production.
Figure~\ref{fig:enter-labelG}(c) shows a comparison of the directional efficiencies of the telescope optics between the design upgrade from nine segments to four segments.
The 4-segment upgrade shows nearly identical efficiencies as the original 9-segment design with a maximum local difference of less than \SI{3}{\percent}.

\begin{figure}[ht]
    \centering
    \subfigure[A PMT measurement setup]{\includegraphics[width=0.25\textwidth]{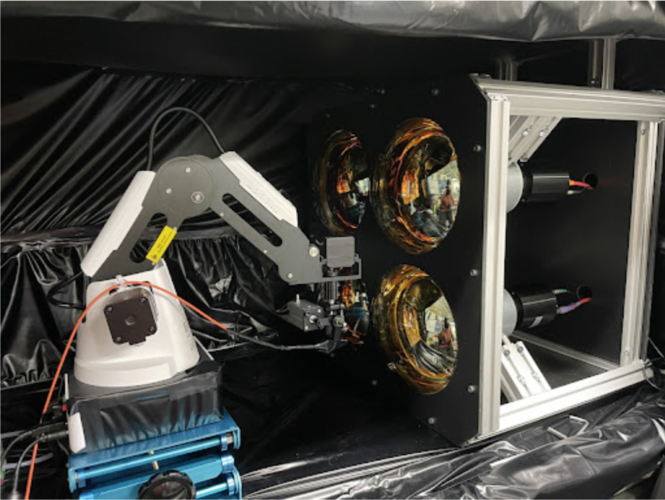}}
    \subfigure[Relative uniformity of the new PMT (R14866)]{\includegraphics[width=0.3\textwidth]{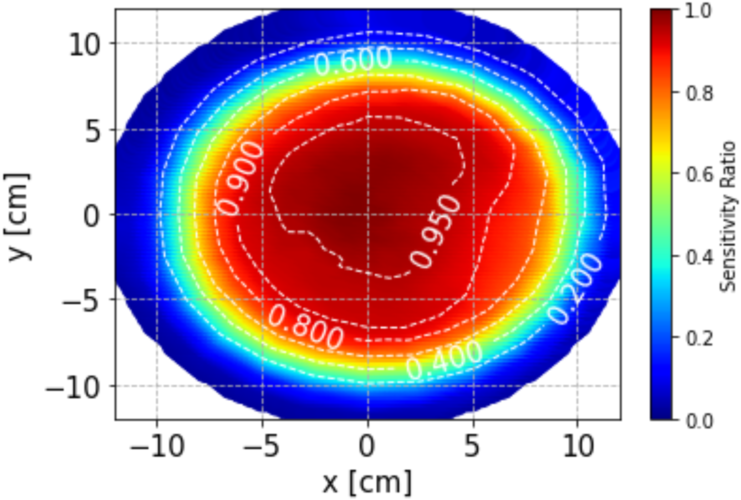}}
    \subfigure[Relative uniformity of the old PMT (R5912)]{\includegraphics[width=0.3\textwidth]{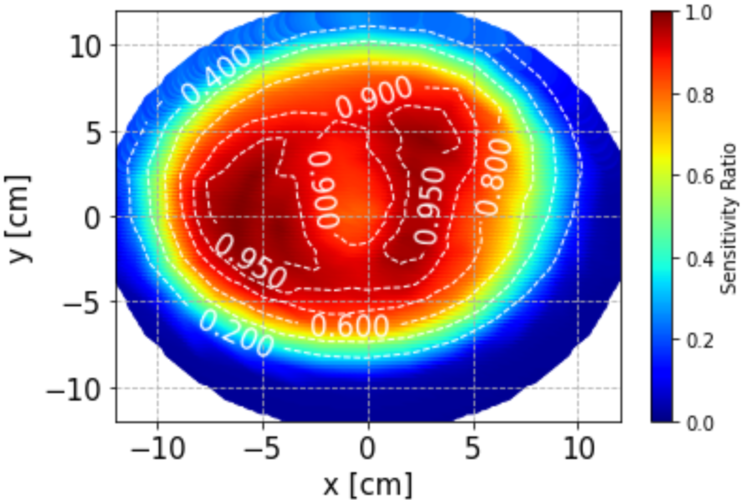}}
    \caption{Uniformity measurements for FAST PMTs. Left: A measurement setup at our laboratory, Middle: Relative uniformity of R14866, Right: Relative uniformity of R5912}
    \label{fig:enter-labelH}
\end{figure}

Calibrations and developments of the PMTs are being performed.
The new PMTs (R14866) are manufactured by Hamamatsu K.K., and have better uniformity in the angular photo detection efficiency than the old design (R5912).
Figure~\ref{fig:enter-labelH} shows the measurement setup and results of uniformity evaluation using the new and old PMTs.
Upgrading the design of the new PMT improved the uniformity by \SI{20}{\percent} on average.
The temperature dependence of the PMT gain is also measured for the calibration.

\section{Summary}
The detection of UHECRs with an order of magnitude larger statistics is essential for shedding further light on their origins and properties.
FAST is one of the next generation UHECR experiments using the atmospheric fluorescence technique.
FAST prototypes are in stable operations in both hemispheres.
Performance estimations with the collected data, with neural network event reconstruction techniques are on going.
Stand-alone observations with autonomous telescopes are essential for the full-scale array.
Further developments and optimizations of the telescope design are in progress.
In parallel, we are focusing on constructing the FAST mini array to validate our observational technique.
Additional prototypes will be installed in coming years.

\section*{Acknowledgements}
This work was supported by JSPS KAKENHI Grant Number 18KK0381, 18H01225, 15H05443, and Grant-in-Aid for JSPS Research Fellow 16J04564 and JSPS Fellowships H25-339, H28-4564. This work was partially carried out by the joint research program of the Institute for Cosmic Ray Research (ICRR) at the University of Tokyo. This work was supported in part by NSF grant PHY-1713764, PHY-1412261 and by the Kavli Institute for Cosmological Physics at the University of Chicago through grant NSF PHY-1125897 and an endowment from the Kavli Foundation and its founder Fred Kavli. The Czech authors gratefully acknowledge the support of the Ministry of Education, Youth and Sports of the Czech Republic project No. LTAUSA 17078, CZ.02.1.01/0.0/17\_049/0008422, CZ.02.1.01/0.0/0.0/16\_013/0001403, LM2023032 and the support of the Czech Academy of Sciences and Japan Society for the Promotion of Science within the bilateral joint research project with Osaka Metropolitan University (Mobility Plus project JSPS 21-10). The Australian authors acknowledge the support of the Australian Research Council, through Discovery Project DP150101622. The authors thank the Pierre Auger and Telescope Array Collaborations for providing logistic support and part of the instrumentation to perform the FAST prototype measurement and for productive discussions.

\bibliography{skeleton}
\bibliographystyle{JHEP}

\end{document}